\documentstyle[prl,aps,preprint,epsf,floats]{revtex}

\begin{document}
\newcommand{\be}{\begin{equation}}
\newcommand{\ee}{\end{equation}}
\newcommand{\bq}{\begin{eqnarray}}
\newcommand{\eq}{\end{eqnarray}}
\newcommand{\Sc}{Schr\"odinger\,\,}
\newcommand{\Sp}{\,\,\,\,\,\,\,\,\,\,\,\,\,}
\newcommand{\Spp}{\,\,\,\,\,}
\newcommand{\no}{\nonumber\\}
\newcommand{\tr}{\text{tr}}
\newcommand{\p}{\partial}
\newcommand{\la}{\lambda}
\newcommand{\La}{\Lambda}
\newcommand{\G}{{\cal G}}
\newcommand{\Gam}{\Gamma}
\newcommand{\GG}{G_s^{i \bar i a \bar a}(x,y)}
\newcommand{\D}{{\cal D}}
\newcommand{\W}{{\bf W}}
\newcommand{\de}{\delta}
\newcommand{\al}{\alpha}
\newcommand{\bi}{\beta}
\newcommand{\ga}{\gamma}
\newcommand{\ep}{\epsilon}
\newcommand{\E}{{\cal E}}
\newcommand{\vep}{\varepsilon}
\newcommand{\th}{\theta}
\newcommand{\om}{\omega}
\newcommand{\J}{{\cal J}}
\newcommand{\pr}{\prime}
\newcommand{\ka}{\kappa}
\newcommand{\TH}{\mbox{\boldmath${\theta}$}}
\newcommand{\DE}{\mbox{\boldmath${\delta}$}}
\newcommand{\ar}{\begin{array}{c}  \\  \Large{ \longrightarrow} \\  ^{s\to 0}\,\end{array}}

\setcounter{page}{0}
\def\footnoterule{\kern-3pt \hrule width\hsize \kern3pt}
\tighten
\title{
Fast Lane for Confinement
}
\author{Jiannis Pachos
\footnote{Email address: {\tt pachos@ctp.mit.edu}}
}
\address{Center for Theoretical Physics \\
Massachusetts Institute of Technology \\
Cambridge, Massachusetts 02139 \\
{~}}

\date{MIT-CTP-2762, ~~July 1998}
\maketitle

\thispagestyle{empty}

\begin{abstract}

Within the Electric Schr\"odinger Representation of the Yang-Mills theory the Hamiltonian eigenstate and eigenvalue, as well as the Coulomb and confining potentials are presented for a special regularization-approximation scheme, which focuses on the ultra-local behavior of the propagator.

\end{abstract}

\vspace*{\fill}

\newpage

\section{Introduction}

In quantum mechanics the physics of the bound states is well distinctive from the physics of the asymptotically free ones. The way we perceive the quantum character for almost free particles, which ``are averaged along all possible paths'' is different from the concept of the ``discrete states'', for example in the hydrogen atom. In quantum field theory the main tool of computations has been the perturbation theory of asymptotically free particles, which gives little insight in the behavior of bound states. On the other hand, the success of the Schr\"odinger representation in describing non-perturbative quantum mechanics suggests its application to field theory for the study of confinement. For a presentation of the field theoretical \Sc representation, see e.g. \cite{Sym1,lusch,Sym2,Sym3,Jack,Paul,MSP,Green,Zarembo}.

Here is shown how the Yang-Mills theory can be studied successfully in the electric (momentum) representation \cite{Jack1,Freedman,Jiannis1,Jiannis,AJ,Jackiw}. This approach provides a natural way to visualize the $SU(N)$ gauge symmetry of the theory. Due to the vector character of the field variables, $E$, in contrast to the connection character of the vector potential, $A$, it is simple to construct gauge invariant quantities. The Gauss' law constraint, which accompanies the Hamiltonian, describes the transformation properties of the wave functional, and also reveals additional constraints on the electric field variable, $E$. They are imposed via delta functions in the wave functional and {\it generate the interactions of the theory}. In order to visualize the behavior of the field, a quark and an anti-quark have been introduced \cite{Rossi}. With simple technical steps the Schr\"odinger equation is shown to be satisfied by a proposed Gaussian wave functional, with ultra-local propagation. This ansatz will not influence the interactions between the sources as they are determined from the constraints on $E$. Though, it will be necessary to restore the propagation in a natural way in order to solve the mass gap equation. The divergencies are regularized by the kernel point splitting method and the renormalization of the theory is performed with proper counter-terms added to the Hamiltonian. In addition the Coulomb and the confining potentials between the static quarks are derived. Their construction is possible in the region where the cut-off is kept finite. The simple structure of this wave functional as well as its similarities with the ground state of the free field suggest that this is the vacuum wave functional of the Yang-Mills theory. 

In the Appendix the free massive scalar field is studied, so that the basic concepts of the momentum representation and of the regularization scheme adopted here are presented pedagogically.

\section{Yang-Mills in the Electric Representation}

The Yang-Mills theory can be formulated in the electric (momentum) representation. Classically the field variable is the vector potential, $A^{i}(x)$, which is element of the $su(N)$ algebra. The magnetic field can be derived from it as $B^{ia}(x)=\ep^{ijk}(\p_jA^{ka}(x)+1/2f^{abc}A^{jb}(x)$ $A^{kc}(x))$, while the electric field, $E^i$, is its conjugate momentum. Although $E^i$ and $A^i$ have both vector transformation properties under a coordinate transformation, they behave differently   under a gauge transformation, $U \! \in \! SU(N)$. $E^i$ transforms as a vector, $E^i\rightarrow {E^i}^U=UE^iU^{-1}$, while $A^i$ as a ``connection'', $A^i \rightarrow {A^i}^U=UA^iU^{-1}-iU\p^i U^{-1}$. The classical Hamiltonian is given by
\be
H=\int d^3 x \Big({g^2 \over 2} E^{ia}(x) E^{ia}(x) + {1 \over 2 g^2}B^{ia}(x)B^{ia}(x) \Big)
\ee
and it is accompanied by the Gauss' law constraint
\be
\p_i E^{ia}(x)+if_{abc} A^{ib}(x) E^{ic}(x)=0 \,\, .
\ee
Here, $i$ runs from $1$ to $3$ when the Hamiltonian is evoked and through $d$ spacial dimensions otherwise. The index $a$ runs over the $N^2-1$ independent components of the $su(N)$ algebra. When the theory is quantized in the electric Schr\"odinger representation the commutation relation
\be
[A^{ia}(x),E^{jb}(y)]=-i \de^{ij} \de^{ab}\de^d(x-y)
\ee
is satisfied for $E^i$ diagonalized and for $A^i$ being a differential operator with respect to $E^i$, $A^{ia}(x)=i{\de \over \de E^{ia}(x)}$. A state in the theory is described by a functional, $\Psi[E]$, which has formally to satisfy the Hamiltonian eigenvalue equation $\hat H \Psi[E]=\E \Psi[E]$ as well as the Gauss' law
\be
G_a(x)\Psi[E] \equiv \left(\p _iE_a^i(x)-if_{abc}E_b^i(x) {\de  \over \de E_c^i(x)}\right)\Psi[E]=0 \,\, .
\label{gauss}
\ee
In the case where static sources are inserted in the theory the Gauss' law is modified to be
\be
G_a(x)\Psi[E]=\Psi[E]T_a \de^d(x-x_0)-T_a\Psi[E] \de^d(x-x_1) \,\, ,
\label{gauss1}
\ee
denoting that there is an $SU(N)$ color source at the point $x_0$ and an anti-source at the point $x_1$. Here, constraint (\ref{gauss1}) will be considered, as the redaction to (\ref{gauss}) is straightforward by taking the points $x_0$ and $x_1$ to infinity. 

\section{Gauss' Law}

Before turning to the \Sc equation the Gauss' law constraint will be studied and solved. The conditions it imposes on the wave functional are crucial to determine the interactions of the theory \cite{papu}. The Gauss' law enforces the following transformation property to the wave functional 
\be
\Psi[E^U]=e^{-{1 \over c}\tr \int E^i \p _iU^{-1} U} U(x_1)\Psi[E] U^{-1}(x_0) \,\, .
\ee
Consequently, its solution is given by \cite{Jiannis1,Jackiw1,Freed1,Nair}
\be
\Psi[E]=\int \D u e^{-{1 \over c} \tr \int E^i \p_iu u ^{-1}} u(x_1)u^{-1}(x_0) \,\,\Phi[E]
\label{fun1}
\ee
The phase factor of the integrand gives the phase of the transformed functional and the group elements defined at the points $x_0$ and $x_1$ give the proper behavior required from the sources. The undetermined functional, $\Phi[E]$, is invariant under gauge transformations. In contrast to the $A$-representation, where gauge invariant objects made out of $A$ are non-local, for example the trace of the Wilson loop, in the $E$-representation the gauge invariant objects are more elementary. They are traces of products of the field $E$ at the same spatial point. 

It is very essential to separate the gauge degrees of freedom from the dynamical variables. It can be effectively done with a natural reparameterization of the electric field, as $E^i(x)=g(x)K^i(x)g^{-1}(x)$, where $g(x) \in SU(N)$ and transforms as $g(x) \rightarrow g^U(x)=U(x)g(x)$, while the $K^i(x)$'s are elements of the corresponding algebra and they do not transform under gauge transformations \cite{Jiannis}. Hence, the $SU(N)$ property of the electric field is clearly separated. In this way a fixed direction in the $SU(N)$ space is defined by $K$ and any other direction required for $E$ can be derived with a $g$ rotation of $K$. This reparameterization needs additional constraints in order to become one-to-one. To define conveniently which parameters to constraint and which not it is important to study how $g$ is constructed. 

A general group element, $g$, can be taken to satisfy
\be
{\cal J}_a g\equiv L_{ab}{\p \over \p \chi^b}g=-T^a g \,\,\,\,\, , \,\,\,\,\,\,\,\,\,\,\,\,\,\,\,\,\,\, {\cal J}^R_a g \equiv R_{ab}{\p \over \p \chi^b}g = g T^a \,\, ,
\label{diag}
\ee
for a special reparameterization, $g(\chi(x))$, with respect to the $N^2-1$ angles, $\chi_a(x)$, and for some invertible matrices $L$ and $R$ \cite{Jiannis}. From the $N^2-1$ matrix hermitian generators, $T^a$, of the $SU(N)$ group, satisfying $\tr T^a T^b =c \de^{ab}$, it is convenient to let the Cartan elements to be the $N-1$ first ones, i.e. $(T^\ka)_{\al \bi}\equiv f^\ka(\al) \de_{\al \bi}$ for $\ka=1..N-1$. Along with this matrix representation of the generators the differential generators ${\cal J}_a$ or ${\cal J}_a^R$ can be organized similarly. The first $N-1$ can be taken to be the differentiation with respect to a single angle, named $\phi^\ka$ for the ``left'' generators and $\bar \phi^\ka$ for the ``right'' ones. Then the first $N-1$ of the relations (\ref{diag}) become
\be
{\cal J}_\ka g\equiv -i{\p \over \p \phi^\ka}g=-T^\ka g \,\,\,\,\, , \,\,\,\,\,\,\,\,\,\,\,\,\,\,\,\,\,\, {\cal J}^R_\ka g \equiv -i{\p \over \p \bar \phi^\ka}g = g T^\ka \,\, ,\, \Sp \ka=1..N-1\,\,,
\ee
which can be solved to find the $\phi$ and $\bar \phi$ dependence of $g$, as 
\be
g=h(\phi)\tilde g(\th) h(\bar \phi)
\ee
where $\theta$ are the remaining $(N-1)^2$ angles of $\chi$, and the $h$ elements belong to the Cartan subgroup, $H$, of $G=SU(N)$, while $\tilde g\in G/H$. 

With the above diagonalization of the $SU(N)$ group it is possible to perform the group integration in relation (\ref{fun1}). The functional turns out to have only exponential dependence on $\bar \phi$, so the $\D \bar \phi$ integration produces a delta function. This makes the $\D \theta$ and $\D \phi$ integrations easy to perform, resulting finally to
\bq
\Psi[E]&=&e^{- {1 \over c} \tr \int E^i \p_ig g^{-1}} \sum_\rho  g(x_1) P^\rho g^{-1} (x_0) 
\no \no
&&
\times \prod_{\ka=1}^{N-1} \prod_x \de \Big( \p_i K^{i\ka}-f^\ka(\rho)\de^d (x-x_0) +f^\ka(\rho) \de^d (x-x_1) \Big) \,\, \Phi[E] \,\, ,
\label{fun2}
\eq
where $P^\rho =\text{diag}(0..0,1,0..0)$, with $1$ in the $\rho$-th place. The first two terms of (\ref{fun2}) give the proper transformation properties of the wave functional. In the case where there are no sources, the constraints in the delta functions are $ \p_i K^{i\ka}=0$ for $\ka=1..N-1$. It is natural then, to extend the constraints to $\p_i K^{ia}=0$ for $a=0..N^2-1$ so that spatial symmetry is restored and the decomposition $E^i=gK^ig^{-1}$ becomes one-to-one. This symmetry is obviously broken when the sources are present.

The constraints are imposed via delta functions and they should be applied at the end of the calculations on the eigenvalue of the operator acting on the wave functional. In this way the functional differentiations with respect to $K$ can be performed as usual and then the constraints can be applied. Also, the functional delta functions approach unity rather than infinity when their argument approaches zero. This is because the integrations which produced them are along the region of the according angles, rather than the region from minus infinity to plus infinity. The extension is made using periodicity, so a normalization factor has to be included, which gives the above property.

The constraints on $K^i$ in the free version of (\ref{fun2}) can also be obtained with a more intuitive consideration. The transformation properties of the wave functional, $\Psi[E^U]=\exp {1 \over c} \tr \int d^d x E^i U^{-1}\p_iU \Psi[E]$, can be satisfied by $\Psi[E]=\exp i\Omega[E] \Phi[E]$, where the phase factor produces the transformation factor, and $\Phi[E]$ is invariant, i.e.
\be
\Omega[E^U]-\Omega[E]=-i {1 \over c} \tr \int d^d x E^i U^{-1} \p_i U \,\, .
\ee
Note that this is the behavior of the phase factor in (\ref{fun2}). Infinitesimally it gives
\be
\Omega [E+[\theta,E]] -\Omega[E]=\int d^d x \p_i E^{ia} \theta^a \,\, ,
\label{phase}
\ee
where $\theta$ is a parameter belonging in the $su(N)$ algebra. If it is chosen such that it commutes with $E$, then the left hand side is zero, while the right hand side need not. This could be resolved by allowing $\Omega[E]$ to have singularities \cite{Jackiw1,Freed1}. Requiring that $\Omega[E]$ is non-singular it is necessary to impose classical constrains on $E$ in the following way. For a specific direction of the electric field, i.e. $K^i$, which lies on the Cartan sub-algebra, the elements $g$ which belong to the Cartan subgroup will commute with it, $[K^i_{Cartan},\theta_{Cartan}]=0$. Then (\ref{phase}) implies that 
\be
\int d^d x \p_i K^{i\ka} \theta^{\ka}=0
\ee
for various configurations of $\theta^\ka$. This is satisfied identically if $\p_iK^{i\ka}=0$ for $\ka=1..N-1$. These are the same constraints, which could be derived straightforwardly with the group integration $\D u$ of the original functional. 

The delta functions in expression (\ref{fun2}) force $K^{i\ka}(x)$ to be the $x_i$ derivative of the Green's function, $F(x)$, which satisfy the following equation
\be
\p_i \p^i F(x) =f^\ka(\rho)\de^d(x-x_0)-f^\ka(\rho)\de^d(x-x_1)
\ee
In $d$ spatial dimensions $K^{i\ka}$ is given by
\be
K^{i\ka} (x)= \Sp \left\{
\begin{array}{ll}
\Sp f^\ka(\rho)[\th(x-x_0)-\th(x-x_1)]& \Sp \mbox{for $d=1$,} \\ \\
\Sp {f^\ka(\rho) \over 2 \pi} \p_i \left(\log {|x-x_0| \over |x-x_1|} \right) & \Sp \mbox{for $d=2$,} \\ \\
\Sp-{f^\ka(\rho) \over (d-2)\omega_d} \p_i \Big({1 \over |x-x_0|^{d-2}}-{1 \over |x-x_1|^{d-2}} \Big) & \Sp \mbox{for $d>2$,}
\end{array}
\right.
\label{green}
\ee
where $\omega_d=2\pi^{d/2}/\Gamma (d/2)$ \cite{Jiannis1}. Apart from the Cartan components the rest ones are also defined uniquely from their constraint. Requiring that $K^{i\eta}(x)$ is finite for any $x$ the following holds
\be
\p_iK^{i\eta}(x)=0 \Rightarrow K^{i\eta}(x)=0 \,\, ,\,\,\,\,\, \text{for} \,\,\,\, \eta=N..N^2-1 \,\, .
\ee
The divergence of $K^{i\ka}(x)$ for $x=x_0$ or $x=x_1$ is due to the sources and does not contradict the finiteness of the field throughout space. The complete determination of the field, $K$, at a first glance is in contrast to the spirit adopted in the path integral consideration where the field is integrated over all possible configurations. The result here is an exact effect coming from the complete separation of the gauge symmetry from the dynamical variables. These values of $K$ generate the interactions, as it is seen in the following. Here, only the longitudinal degrees of freedom are considered for the solution of the Abelian-like Gauss' law, as this is also the part which generates the interactions in the Electromagnetism. It has been argued that the transverse part is related to dual monopole interactions \cite{Shara}. Even though we do not study this part its possible contribution would be an additional effect to the results presented here.

\section{Hamiltonian}

The successful separation of the gauge degrees of freedom from the dynamical ones makes the gauge fixing procedure trivial. It is necessary to fix the symmetry of the rotations with respect to $g \in SU(N)$, as these variables will cause extra divergences when the functional differentiations are taken \cite{Jack1,Freedman}. These divergences are not important in the dynamics of the theory because the physical variables have to be independent from the gauge parameter $g$. The divergences are similar to the ones which make the gauge fixing necessary when the equivalent path integral framework is used. But now rather than needing to insert ghost fields, the gauge fixing can be performed naturally by taking $g$ to be constant throughout space, e.g. equal to the identity rotation. 

In the gauge $g= {\bf 1}$, the wave functional becomes
\be
\Psi[K]=\sum_\rho  P^\rho \prod_{\ka=1}^{N-1} \prod_x \de \Big( \p_i K^{i\ka}-f^\ka(\rho)\de^3 (x-x_0) +f^\ka(\rho) \de^3 (x-x_1) \Big) \,\, \Phi[K] \,\, .
\label{fun3}
\ee
The presence of the sources is still denoted with the delta functions. The unknown part is the gauge invariant functional $\Phi[K]$, which will be determined from the \Sc equation. With the $SU(N)$ rotations fixed towards the $K^i$ direction the \Sc equation becomes
\be
\hat H[K] \Psi[K] \equiv \int d^3 x \Big( { g^2 \over 2} K^{ia}(x)K^{ia}(x) +{1 \over 2 g^2} B^{ia}(K(x)) B^{ia}(K(x))\Big)\Psi[K]= \E \Psi[K]
\label{ham1}
\ee
with
\be
B^{ia}(K(x))=i \ep ^{ijk} \Big( \p_j { \de \over \de K^{ka}(x)} + {i \over 2 } f^{abc} { \de \over \de K^{jb}(x)} { \de \over \de K^{kc}(x)} \Big) \,\, .
\label{magn1}
\ee
In order to proceed with the functional differentiations it is convenient to make an ansatz for $\Phi[K]$. It will be taken to be, as in the free case with ultra-local propagation (see Appendix), a function of the invariant quadratic form $M[K]\equiv 1/\La \int d^3x K^{ia}(x) K^{ia}(x)$. The function dependence of $\Phi$ on the quadratic form, $M[K]$, will be taken for the ground state to be an exponential, as it is the simplest possible functional which indeed satisfies (\ref{ham1}) within the approximate scheme presented in the Appendix. In order $M[K]$ to be dimensionless, $\La$ has dimensions of [mass] or $1/$[length]. 

Applying the magnetic operator on $\Phi(M)$ it is obtained
\be
B^{ia}(K(x))\Phi(M)=i\ep^{ijk} \Big( { 2 \over \La} \p_jK^{ka}(x) \Phi^{\pr}(M)+{i \over 2} ( {2 \over \La} )^2 f^{abc} K^{jb}(x)K^{kc}(x) \Phi^{\pr \pr} (M) \Big) \,\, ,
\ee
where the primes denote derivatives with respect to $M$. The action of the two functional derivatives of the magnetic field (\ref{magn1}) on the functional, gives a divergence $\de^3 (0)$, but it drops out after the contraction of two indices of the antisymmetric structure constants, $f^{abc}$. Hence, there is no need to introduce regularization for defining the action of the magnetic field on the functional. This result is connected with taking the propagator in $M[K]$ to be ultra-local. Though, the action of the two magnetic fields of the Hamiltonian on the functional will produce divergences, so a regularization scheme is necessary. For this the point splitting method is adopted with the help of a kernel, $G^{i \bar i a \bar a}_s(x,y)$, which has the following properties
\be
G_s^{i \bar i a \bar a}(x,y) \ar \de^3(x-y) \de^{i \bar i} \de^{a \bar a} \Sp \text{and} \Sp G^{i \bar i a \bar a}_s(x,x)=\G_s(0)\de^{i \bar i} \de^{a \bar a}
\ee
where $\G_s(x,y)$ is a regularization of the delta function, for $s \rightarrow 0$. As the $SU(N)$ geometry of the theory has been gauged out, the kernel $G_s^{i \bar i a \bar a}(x,y)$ does not have to satisfy any group transformation properties. So $G_s^{i \bar i a \bar a}(x,y)$ can be taken to be
\be
G_s^{i \bar i a \bar a}(x,y)=\G_s(x,y) \de^{i \bar i} \de^{a \bar a}
\label{ker1}
\ee
without any loss of generality. The action of the regularized magnetic term of the Hamiltonian on $\Phi(M)$ gives
\bq
{1 \over 2 g^2} \int d^3x d^3y \GG && B^{\bar i \bar a}(K(y)) B^{ia}(K(x)) \Phi(M)=
\no \no
- {1 \over 2 g^2} \int d^3x && d^3y \ep^{ijk} \ep^{\bar i \bar j \bar k} \Bigg[ {\p}^y_{\bar j} {\p}^x _j \GG { \de \over \de K^{\bar k \bar a} (y)}{ \de \over \de K^{k a} (x)} \Phi(M)-
\no \no
&&{i \over 2} \p^y _{\bar j} \GG f^{abc} {\de \over \de K^{\bar k \bar a}(y)}{\de^2 \over \de K^{jb}(x) \de K^{ka}(x)} \Phi(M) -
\no \no
&&{i \over 2} \p^x_j \GG f^{\bar a \bar b \bar c} {\de^2 \over \de K^{\bar j \bar b}(y) \de K^{\bar k \bar a}(y)} {\de \over \de K^{ka}(x)} \Phi(M) -
\no \no
&&{1 \over 4} \GG f^{abc}f^{\bar a \bar b \bar c} {\de^2 \over \de K^{\bar j \bar b}(y) \de K^{\bar k \bar a}(y)}{\de^2 \over \de K^{jb}(x) \de K^{ka}(x)} \Phi(M) \Bigg] \,\, .
\eq
The two functional derivatives acting on $\Phi(M)$ give
\be
{ \de \over \de K^{\bar k \bar a} (y)}{ \de \over \de K^{k a} (x)} \Phi(M)= {2 \over \La} \de^{k \bar k} \de^{a \bar a} \de^3(x-y) \Phi'(M)+({2 \over \La})^2 K^{\bar k \bar a}(y) K^{ka}(x) \Phi''(M)
\ee
while the three derivatives contracted with $f^{abc}$ give
\bq
f^{abc}{\de \over \de K^{\bar k \bar a}(y)}{\de^2 \over \de K^{jb}(x) \de K^{kc}(x)}&& \Phi(M)=
\no \no
f^{abc}({2\over \La})^2\Big(K^{jb}(x)\de^{\bar k k}&&\de ^{\bar a c}\de^3 (x-y) + K^{kc}(x) \de ^{\bar k j} \de^{\bar a b} \de^3(x-y)\Big) \Phi''(M)+
\no \no
f^{abc}&&({2 \over \La})^3 K^{kc}(x)K^{jb}(x)K^{\bar k \bar a}(y) \Phi'''(M)
\eq
as well as
\bq
f^{\bar a \bar b \bar c} {\de^2 \over \de K^{\bar j \bar b}(y) \de K^{\bar k \bar c}(y)}{\de \over \de K^{ka}(x)}&& \Phi(M)=
\no \no
f^{\bar a \bar b \bar c}({2\over \La})^2\Big(K^{\bar j \bar b}(y)\de^{\bar k k}&&\de ^{\bar c a}\de^3 (x-y) + K^{\bar k \bar c}(y) \de ^{\bar j k} \de^{\bar b a} \de^3(x-y)\Big) \Phi''(M)+
\no \no
f^{\bar a \bar b \bar c}&&({2 \over \La})^3 K^{ka}(x)K^{\bar k \bar c}(y) K^{\bar j \bar b}(y) \Phi'''(M) \,\, .
\eq
The term with four functional derivatives gives
\bq
f^{abc} f^{\bar a \bar b \bar c} { \de^2 \over \de K^{\bar j \bar b}(y) \de K^{\bar k \bar c}(y)} {\de^2 \over \de K^{j b}(x) \de K^{kc}(x)} \Phi&&(M)=
\no \no
f^{abc} f^{\bar a \bar b \bar c} \Big[ ({2 \over \La})^2 \Big( \de^{\bar k k} \de^{\bar c c} \de^{\bar j j} \de^{\bar b b}+  \de^{\bar j k} \de^{\bar b c}\de^{\bar k j} \de^{\bar c b} && \Big) (\de^3(x-y))^2 \Phi''(M)+
\no \no
({ 2 \over \La})^3 \Big( K^{\bar j \bar b}(y) K^{jb}(x)\de^{\bar k k} \de^{\bar c c} +&& K^{\bar j \bar b}(y) K^{kc}(x) \de^{\bar k j} \de^{\bar c b} + 
\no \no 
K^{\bar k \bar c}(y) K^{jb}(x) \de^{\bar j k} \de^{\bar b c} +&& K^{\bar k \bar c} (y) K^{kc}(x) \de^{\bar j j} \de^{\bar b b} \Big) \de^3(x-y) \Phi'''(M)+
\no \no
&&
({2 \over \La})^4 K^{\bar j \bar b}(y) K^{\bar k \bar c}(y) K^{jb}(x) K^{kc}(x) \Phi''''(M) \Big]
\eq
Substituting these expressions in the regularized Hamiltonian acting on $\Phi(M)$, we finally obtain
\bq
H_s \Phi(M)=&&{ g^2 \over 2} \int d^3x K^{ia}(x)K^{ia}(x) \Phi(M) -
\no \no
{1 \over 2 g^2}&& \int d^3 x d^3y \ep^{ijk} \ep^{\bar i \bar j \bar k}  \Bigg[({2 \over \La}) \p^y_{\bar j} \p^x_j \GG \de^{\bar a a} \de^{\bar k k} \de^3(x-y) \Phi'(M) -
\no \no
&&
2 {1 \over 4} ({ 2 \over \La})^2 f^{abc} f^{\bar a \bar b \bar c} \GG \G_s(x,y) \de^3(x-y) \de^{\bar j j} \de^{\bar b b} \de ^{\bar k k} \de^{\bar c c} \Phi '' (M)+
\no \no
&&
2 { i \over 2} ({ 2 \over \La})^2 \de ^3(x-y) \Big( \p^x_j G_s^{\bar i i \bar a a }(y,x)+ \p^x_j \GG \Big) \de^{\bar k k} f^{a \bar b \bar a} K^{\bar j\bar b}(x) \Phi'''(M)+
\no \no
&&
({2 \over \La})^2 \p^y_{\bar j} \p^x_j \GG K^{\bar k \bar a}(y) K^{ka}(x)\Phi''(M)-
\no \no
&&
4 {1 \over 4} ({ 2 \over \La})^3 \GG  \de^3 (x-y) f^{abc} f^{\bar a \bar b \bar c} \de^{\bar k k} \de^{\bar c c} K^{\bar j \bar b}(y) K^{j b}(x) \Phi'''(M) -
\no \no
&&
{i \over 2}({ 2 \over \La})^3\Big( \p^x_j G_s^{\bar i i \bar a a }(y,x)+ \p^x_j \GG \Big) f^{\bar a \bar b \bar c} K^{ka}(x) K^{\bar j \bar b}(y) K^{\bar k \bar c}(y) \Phi''' (M)-
\no \no
&&
{1 \over 4} ({2 \over \La})^4 f^{abc} f^{\bar a \bar b \bar c} \GG K^{jb}(x) K^{kc}(x) K^{\bar j \bar b}(y) K^{\bar k \bar c}(y) \Phi ''''(M) \Bigg]
\eq
where the square of the delta function, $(\de^3(x-y))^2$, coming from the action of four derivatives has been regularized to be $\G_s(x-y)\de^3(x-y)$. The term linear in $K$ vanishes for 
\be
\Big( \p^x_j G_s^{\bar i i \bar a a}(y,x)+ \p^x_j \GG \Big)=0
\ee
which reveals the oddness of the first derivative of the kernel, $G$. In addition the terms cubic and quartic in $K$ vanish for the following reason; the field, $K^{ia}(x)$, is constrained from the functional delta functions to be $K^{ia}(x)=\p_i g^a(x)$, for an appropriate function $g^a(x)$. For this value of $K$ the contraction with the antisymmetric $f^{abc}$ gives zero. 

Substituting the exponential dependence of $\Phi$ on $M[K]$ we eventually obtain
\bq
H_s &&\Phi(M)=
\no \no
&&
{1 \over 2g^2} [\int d^3x] \Big( -( {2 \over \La}) \ep^{ijk} \ep^{\bar i \bar j k} \left. \p^y_{\bar j} \p ^x_j G_s^{i \bar i a a} (x,y)\right|_{y=x} + ({2 \over \La})^2 3(N^2-1) C_s(N) \G_s(0)\Big) \Phi(M) +
\no \no
\int && d^3x d^3y \Bigg[ - {1 \over 2g^2} ({2 \over \La})^2 \ep^{ijk} \ep^{\bar i \bar j \bar k} \p^y_{\bar j} \p^x_j G_s^{i \bar i a \bar a}(x,y)+
\no \no
&&
\Sp \Big( {g^2 \over 2} \de^{k \bar k} \de^{a \bar a} + {1 \over 2 g^2} ({2 \over \La})^3 2 C_2(N) G_s^{k \bar k a \bar a} (x,y)\Big) \de^3(x-y)\Bigg]K^{ka}(x) K^{\bar k \bar a}(y)  \Phi(M) \,\, .
\label{ham3}
\eq
This expression is similar to the equivalent expression for the free massive scalar field. We can demand that for $s \rightarrow 0$, then $H_s\Phi(M)=\E_s \Phi(M)$. The term quadratic in $K$ has to vanish and this will determine the value of $\La$. This value will, by its turn, determine the eigenvalue of the Hamiltonian, which is the first part in the RHS of equation (\ref{ham3}). Regularizing the delta function by substituting it with $\G_s(x,y)$ the following equation has to hold
\be
g^4 ({ \La \over 2})^2 \de^{k \bar k} \de^{a \bar a}\G_s(x,y)= \ep^{ijk} \ep^{\bar i \bar j \bar k}\p^y_{\bar j} \p^x_j \GG - ({2 \over \La}) \ep^{ijk} \ep^{\bar i \bar j \bar k} f^{abc} f^{\bar a \bar b c} \G_s(0) G_s^{i \bar i b \bar b}(x,y) \,\, ,
\label{ker4}
\ee
for small $s$. Using the representation of the kernel given in (\ref{ker1}) and keeping in mind that $\La$ is negative, we obtain
\be
{3 \over 2} g^4 \left|{\La  \over 2}\right|^2 \G_s(x,y) = \Big(- \p^2_x +3\left| { 2 \over \La}\right|C_2(N) \G_s(0) \Big) \G_s(x,y) \,\, .
\label{ker5}
\ee
It is convenient at this point to restore the propagator and its inverse by defining
\be
G^{-1}(x,y)\equiv g^2 \left|{\La  \over 2}\right|\G_s(x,y) \Sp \text{and} \Sp G(x,y)\equiv {1 \over g^2} \left|{2  \over \La}\right|\G_s(x,y)
\ee
and also
\be
G^{-2}(x,y) \equiv \int d^3z G^{-1}(x,z) G^{-1}(z,y) =g^4 \left|{\La  \over 2}\right|^2 \G_s(x,y) \,\, .
\ee
Then (\ref{ker5}) becomes
\be
{3 \over 2} G^{-2}(x,y)=(-\p^2_x+\mu^2)\G_s(x,y)
\label{mass1}
\ee
with $\mu^2=3 g^2 C_2(N) G(x,x)$ being a mass scale. Obviously $\mu^2$ is divergent. It is necessary to subtract this divergence from the initial Hamiltonian by adding to it a proper counter-term. This would be the mass term
\be
H_{c.t.}^{mass}={\tilde \mu_0^2 \over 2} \int d^3 x E^{ia}(x)E^{ia}(x)
\ee
which does not violate the gauge invariance of the theory. As an effect will have to re-enter the mass gap equation (\ref{mass1}) with a finite mass $m^2 \equiv \mu^2-\mu_0^2$, with $\mu_0^2=(3/2)|\La /2|^2 \tilde \mu^2_0$, as
\be
{3 \over 2} G^{-2}(x,y)=(-\p^2_x+m^2)\G_s(x,y) \,\, .
\label{mass2}
\ee
This equation can be easily solved to obtain
\be
G^{-1}(x,y)= \sqrt{{ 2 \over 3}} \int {d^3 p \over (2 \pi)^2} e^{i \vec{p} \cdot (\vec{x} - \vec{y})} \sqrt{p^2 + m^2} \Spp \text{and} \Spp  G(x,y)= \sqrt{{ 3 \over 2}} \int {d^3 p \over (2 \pi)^2} e^{i \vec{p} \cdot (\vec{x} - \vec{y})} {1 \over \sqrt{p^2 + m^2}} \,\, ,
\label{renorm1}
\ee
where the parameter, $s$, can be reinterpreted as a cut-off, $S=1/\sqrt{s}$, for the momentum integrations. Hence, we obtain that the mass parameter, $m^2$, is given by
\be
m^2= -\mu^2_0 + g^2 \Gam  \Big[ S^2 + m^2 \Big( {1 \over 2}-\ln {2 S \over m} \Big) \Big], \Sp \Gam \equiv {1 \over 2 \pi^2} 3\sqrt{{3 \over 2}} C_2(N)  \,\, .
\label{mass3}
\ee
It is possible to rewrite this expression with respect to renormalized quantities for the mass and the coupling constant as follows \cite{Kerman,PahOh}
\be
\mu_R^2 = {- \mu_0^2 +g^2 \Gam S^2 \over 1 +g^2 \Gam \ln {2 S \over M}} \Sp \text{and} \Sp  g_R ^2 =- {g^2 \over 1+ g^2 \Gam \ln {2 S \over M}} \,\, ,
\label{renorm2}
\ee
where $M$ is the scale of the theory. With respect to these finite quantities expression (\ref{mass3}) takes the following form
\be
m^2 =\mu_R^2 - g_R ^2 m^2  \Gam \Big( {1 \over 2} + \ln { m \over M} \Big) \,\, .
\ee
In the definitions (\ref{renorm2}), the renormalized coupling constant, $g_R$, scales with respect to the mass $M$. From this relation it is possible to evaluate the beta function, $\beta(g_R)$, of the theory. It is given from
\be
\beta(g_R) \equiv M { \p g_R\over \p M} = - { g_R^3 \over (2 \pi)^2 } 3 \sqrt{{3 \over 2}} C_2(N) \,\, .
\ee
Note that 
\be
3 \sqrt{{3 \over 2}}=3.674 ... \approx {11 \over 3} =3.666 ...
\ee
which is a surprising feature of the theory, as it includes almost the same numerical factor (eleven over three) with the one you can calculate from the perturbation theory with the use of Feynman diagrams. Though, notice the difference where here we have $(2 \pi)^2$ instead of $(4 \pi)^2$. The beta function is negative as expected denoting the asymptotic freedom of the theory.

The relation for the beta function can be integrated out to give a finite mass scale in terms of the renormalized coupling constant
\be
m^2_0 \equiv M^2 e^{-{2 \over  g^2_R \Gam}} \,\, .
\ee
$m_0^2$ is the dynamically generated mass of the theory. 

In addition the eigenvalue of the energy is given by
\be
\E_S= {1 \over 2} [\int d^3 x] 3 (N^2-1) \Big(G^{-1}(0) -{2 m^2 \over 3} G(0) +g^2 C_2(N) (G(0))^2 \Big) \,\, .
\ee
This relation is similar in structure with the one presented in \cite{Kerman} for the $\phi^4$ theory in $(3+1)$ dimensions calculated with the variational method. The divergences are at most quartic in the momentum cut-off, $S$. We shall not proceed with the subtraction of the infinities here as the ultraviolet behavior of $\E_S$ is already apparent.

We expect to obtain the interactions between the sources from the expectation value of the quadratic term in the field $K^i$ \cite{Jiannis1}. But this term was demanded to vanish for $s \to 0$ in the procedure of solving the \Sc equation. In order to turn up the interactions in this approach, we need to keep $s$ finite different from zero. Neglecting the mass counter-term, the quadratic term becomes
\be
- { \tilde g^2 \over 2}\int d^3 x K^{ia}(x) K^{ia}(x) + { \tilde g^2 \over 2} \int d^3 x d^3 y \G_s(x,y) K^{ia}(x) K^{ia}(y) 
\label{inter}
\ee
where
\be
{ \tilde g^2 \over 2}=- {g^2 \over 2} + {1 \over g^2} \left|{2 \over \La}\right|^3 C_2(N) \G_s(0) 
\ee
We shall use for $\La$ the small $g^2$ expansion (\ref{lala}) given in the Appendix, and actually as a first approximation only the leading term. Then 
\be
{\tilde g ^2 \over 2} =- {g^2 \over 2} + {C_2(N) \over \sqrt{4 \pi}^3} g^4 
\ee
where we keep in mind that the unrenormalized coupling constant tends to zero as inverse logarithm when $s \to 0$. Though, we shall keep $g^2$ large enough so that $\tilde g ^2$ is positive. The delta functions in the wave functional enforce $K^i$ to take the specific values given in (\ref{green}). These determine the form of the interaction terms as follows. The first term in (\ref{inter}) gives the following interaction
\be
-{\tilde g ^2 \over 2} \int d^3 x K^{ia}(x) K^{ia}(x)=-V^0_{se}-V^1_{se} + {\tilde g ^2 \over 2} { f^\ka(\rho) f^\ka(\rho) \over \omega_3} {1 \over 4 \pi} {1 \over |x_1 - x_0|}
\ee
where $V^0_{se}$ and $V^1_{se}$ are the Coulomb self-energies of the sources at the points $x_0$ and $x_1$. A sum of its expectation value over the $N+1$ directions the Cartan sub-algebra could take will result to the complete Coulomb potential
\be
V_{Coul}(|x_1-x_0|)= +{\tilde g^2 \over 2} C_2 (N) {1 \over 4 \pi} {1 \over |x_1 - x_0|} \,\, .
\ee
Note that the plus sign in front of the Coulomb potential is causing a repulsion between the static sources in contrast of what would be obtained if only the electric part of the Hamiltonian was used \cite{Jiannis1}. On the other hand, enforcing the constraints of $K^i$ on the second term will result to
\be
{\tilde g^2 \over 2}\int d^3 x d^3 y \G_s(x,y) K^{ia}(x) K^{ia}(y)= {\tilde g^2 \over 2}\int d^3 x d^3 y (- {\p_i^x} ^2) \G_s(x,y) g^a(x) g^a(y) \,\, ,
\ee
where for small $s$ the derivatives on the kernel give 
\be
-\p^2 \G_s (x,y) =-\Big(4{(x-y)^2 \over (4s)^2} - {3 \over 2s} \Big) \G_s(x,y) \rightarrow { 3 \over 2s} \G_s (x,y) \Sp \text{for $x \approx y$,} 
\ee
where we have neglected the $(x-y)^2$ term. Hence
\bq
&&
{\tilde g^2 \over 2}\int d^3 x d^3 y \G_s(x,y) K^{ia}(x) K^{ia}(y) \rightarrow
\no \no
&&
{\tilde g^2 \over 2}{3 \over 2 s} { f^\ka(\rho) f^\ka(\rho) \over \omega_3} \int d^3 x \Big( {1 \over |x -x_0|} - {1 \over |x -x_1|} \Big)^2 = {\tilde g^2 \over 2} {3 \over 2s} { f^\ka(\rho) f^\ka(\rho) \over \omega_3} {1 \over 4 \pi} |x_1-x_0|
\eq
for $s$ small to re-establish $\G_s(x,y)$ as a delta function, but also big enough to make $\tilde g^2$ positive and result to confinement. Taking the expectation value and summing over all direction of the Cartan sub-algebra the confining potential between the sources is obtained
\be
V_{Conf}(|x_1-x_0|)={\tilde g^2 \over 2} {3 \over 2s} C_2 (N) {1 \over 4 \pi} |x_1-x_0| \,\, .
\ee
The existence of the multiplicative factor $1 \over s$ is expected also from dimensional reasons. As $s$ is kept non-zero, it does not pose any divergency problems. However, the approximation step taking $x \approx y$ can be justified by enforcing non-trivial analyticity on the kernel $\G_s(x,y)$. A similar consideration has been assumed in the literature for the calculation of the confining potential \cite{Diakonov,Kondo}.

\section{Conclusions}

Yang-Mills theory presents with its non-linearity many calculational challenges. Here, it is studied in the \Sc electric representation. An exact separation of the gauge degrees of freedom has determined the dynamics of the theory. In order to make the calculations with the application of the Hamiltonian simpler a Gaussian wave functional has been assumed with ultra-local propagation. The propagator has been re-established in a proper way in order to calculate the mass gap of the theory. A similarity of the beta function with the known one loop result nests on the similarity of our method with the variational approximation with which the one loop perturbative calculations should agree \cite{Jackiw2}. In particular the variational method would not pose any difficulties to be applied within this framework as the $SU(N)$ symmetry has been gauged out of the wave functional. 

In principle, this method can be used also in the $A$ representation. The results from the Gauss' law should be identical as has been argued for the simpler model of Electromagnetism in \cite{Hatfield} within the \Sc representation. 

The flexibility of the \Sc representation has revealed with simple steps the main characteristics and behavior of the Yang-Mills theory. We strongly believe that considerations along these lines can be proven fruitful towards further analytical understanding of confinement.

\appendix

\section{Free Scalar Massive Field} 

The free massive scalar field will be studied in a regularization approximation scheme, in which the non-local propagator can be substituted with a scaled local one. This method gives the same leading ultra-violet behavior for the energy eigenvalue, as it is shown in the following.

The Hamiltonian for the free massive scalar field is given by
\be
H={1 \over 2} \int d^3 x \Big(\pi^2(x) -\p_i\phi(x)\p_i\phi(x)+m^2 \phi^2(x)\Big)
\ee
In the momentum representation the quantization condition, $[\phi(x),\pi(y)]=-i\delta^3(x-y)$, can be satisfied by diagonalizing the field variable $\pi(x)$, while $\phi(x)$ is given by $\phi(x)=$ $i {\de \over \de \pi(x)}$. The Hamiltonian is regularized with a point splitting kernel, $\G_s(x,y)=$ $\exp({-(x-y)^2\over 4s})/\sqrt{4 \pi s}^3$, obtaining
\be
H_s={1 \over 2} \int d^3 x d^3 y\G_s(x,y) \pi(x) \pi(y) +{1 \over 2} \int d^3x d^3y \Big(\p_i^x\p_i^y+m^2\Big)\G_s(x,y){\de^2\over \de\pi(x)\de \pi(y)} \,\, .
\ee
The functional, $\Psi[\pi]$, has to satisfy the equation, $H_s \Psi[\pi]=\E_s \Psi[\pi]$, for $s \rightarrow 0$, where we are interested in the way the energy eigenvalue diverges. With the wave functional of the form
\be
\Psi[\pi]=\exp {1 \over \La}\int d^3z d^3 \bar z \,\, \pi(z) g(z, \bar z) \pi(\bar z) \,\, ,
\ee
where $\La$ is negative for normalizability, we obtain
\bq
H_s \Psi[\pi]= -{1 \over 2} {2 \over \La} \int &d^3x& d^3y (\p ^x_i \p^y_i +m^2)\G_s(x,y)g(x,y) \Psi[\pi]-
\nonumber \\ \no
{1 \over 2} \int d^3 z d^3 \bar z \Big[ \Big({2 \over \La} \Big)^2\int &d^3 x& d^3 y (\p^x_i\p^y_i+m^2) \G_s(x,y) g(x,z)g(\bar z,y) -\G_s(z,\bar z) \Big] \pi(z)\pi(\bar z) \Psi[\pi] \,\, .
\label{mon}
\eq
The term depending quadratically in $\pi$ has to vanish. This will determine $g$ and $\La$, and eventually the energy $\E_s$.
Switching to the momentum representation of the coordinates $\vec{x}$ we can easily find that the coefficient of $\pi(z) \pi( \bar z)$ vanishes for
\be
g(x,y)=\int {d^3p \over (2 \pi)^3} e^{i \vec{p} \cdot (\vec{x}-\vec{y})}{ 1 \over \sqrt{p^2+m^2}} \,\,\,\,\,\,\,\,\,\,\,\,\,\,\,\,\,\,\,\,\,\,\,\, \text{and} \,\,\,\,\,\,\,\,\,\,\,\,\,\,\,\,\,\,\,\,\,\,\,\, \La=-2 \,\, ,
\ee
where the regularization with $\G_s(x,y)$ can be reinterpreted as a cut-off regularization in the momentum integrations. Note, that this is a choice where the propagator of the field is non-local.

There is a way to approximate the propagator with a scaled ultra-local one. To achieve this, $\La$ has to diverge as $s$ goes to zero. For $g(z, \bar z)=\de^3(z-\bar z)$ equation (\ref{mon}), becomes
\bq
H_s \Psi[\pi]=&& -{1 \over 2} {2 \over \La_s} \int d^3x \left. (\p ^x_i \p^y_i +m^2)\G_s(x,y)\right|_{y=x} \Psi[\pi]-
\no \no
&&
-{1 \over 2}\Big({2 \over \La_s} \Big)^2 \int d^3 x d^3 y \Big[ (\p^x_i\p^y_i+m^2) \G_s(x,y) -\Big({\La_s \over 2} \Big)^2 \G_s(x, y) \Big] \pi(x)\pi(y) \Psi[\pi] \,\, .
\label{mon1}
\eq
The vanishing of the quadratic term in $\pi$ becomes exact for $y=x$ as
\be
\left. (-\p^2+m^2)\G_s(x,y)\right|_{y=x}=\Big( {\La_s \over 2} \Big)^2 \left. \G_s(x,y) \right|_{y=x} \Rightarrow \left({3 \over 2s}+m^2\right) \G_s(0)=\Big( {\La_s \over 2} \Big)^2  \G_s(0) \,\, ,
\label{nom2}
\ee
for $(\La_s/2)^2={3 \over 2s} +m^2$. Equation (\ref{nom2}) represents the Lorentz invariance of the theory denoted with the relativistic relation $E^2=p^2+m^2$ in a regularized way. The region $y=x$, is the most dominant for the energy eigenvalue. For this choice of $g(z, \bar z)$ the wave functional $\Psi[\pi]$ has become a pure Gaussian. Applying the Hamiltonian operator on $\Psi[\pi]$ the expected eigenvalue has the form
\be
\E_s=-{1 \over 2} {2 \over \La_s} [\int d^3x]\left. (\p ^x_i \p^y_i +m^2)\G_s(x,y)\right|_{y=x}={1 \over 2}[\int d^3x] \G_s(0) \sqrt{{3 \over 2s} +m^2} \,\, ,
\ee
where the value at $y=x$ is evoked. $\E_s$ has the right $\sim 1 /s^2$ behavior for small $s$, which is a quartic divergence with respect to the momentum cut-off, $S=1/\sqrt{s}$. Note that (\ref{nom2}) holds only approximately for $y \neq x$ as we have truncated the propagator to be an ultra-local one.

The divergence of the wave functional due to $\La_s$ is connected to the energy divergence due to the equal time procedure adopted in the Schr\"odinger representation, and it can be absorbed in the fields of the theory, as has been pointed out by Symanzik \cite{Sym1}. Inserting the renormalization finite scale $\bar s$, the field, $\pi(x)$, can be re-defined as 
\be
\pi(x) \rightarrow \bar \pi (x)\equiv\sqrt{3/(2 \bar s)+m^2 \over 3/(2s)+m^2} \,\,\pi(x)=\sqrt{\La_{\bar s} \over \La_s}\,\, \pi(x) \,\, .
\ee
Then, the wave functional takes the renormalized form 
\be
\Psi[\bar \pi]=\exp {1 \over \La_{\bar s}} \int d^3x \,\, \bar \pi(x) \bar \pi(x)\,\, ,
\ee
while the renormalized energy is given, with a re-definition of its zero point, by
\be
{\E_{\bar s} \over [\int d^3x] \G_s(0)} ={1 \over 2} \sqrt{{3 \over 2 \bar s} +m^2}\,\, .
\label{en2}
\ee
This completes the study of the particular renormalization of the free field.

\section{Evaluation of $\La$ for Yang-Mills}.

It is possible to solve (\ref{ker5}) for $y=x$ with respect to $\La$ as a function of, $s$, by using the kernel given in (\ref{ker1}). Then, this equation becomes in terms of $\La_s$ 
\be
{3 \over 2} \Big( g^2\left| {\La_s \over 2} \right| \Big)^2= {3 \over 2 s} +3 C_2(N)g^2 \Big(g^2\left|{\La_s \over 2 }\right|\Big)^{-1}\G_s(0) \,\, ,
\label{maga}
\ee
where the explicit representation for the regularized delta function is taken to be 
\be
\G_s(x,y)={e^{-{(x-y)^2 \over 4s}} \over \sqrt{4 \pi s}^3}
\ee
This equation is cubic in $\La_s /2$ and can be solved as an expansion for small $g^2$ to be
\be
{\La_s \over 2} =- {1 \over \sqrt{s} g^2} \Big( 1 +{C_2(N) \over \sqrt{4 \pi }^3} g^2 -{ 3 C_2^2(N) \over 128 \pi^3 } g^4 +{C_2^3(N) \over 16 \sqrt{4 \pi}^3 }g^6 + \cdots \Big) \,\, ,
\label{lala}
\ee
where we have chosen the solution with negative values for small $g^2$ for normalizability of the wave functional. For this calculation of $\La_s$ from the mass gap equation (\ref{maga}) the mass counter-term, $H^{mass}_{c.t.}$, has been overlooked. Within this assumption the calculation of the interaction terms has been performed with the use of the above expansion of $\La_s$.

\end{document}